\documentclass{jnmp}
\begin{document}

\renewcommand{\evenhead}{G~Gaeta,~R~Mancinelli}
\renewcommand{\oddhead}{Asymptotic scaling in ARD equations}

\def\giorno{14/2/2005}

\def\*{{\bf ***}}

\def\a{\alpha}
\def\b{\beta}
\def\g{\gamma}
\def\ga{\gamma}
\def\de{\delta}   
\def\eps{\varepsilon}
\def\phi{\varphi}
\def\la{\lambda}
\def\ka{\kappa}
\def\r{\rho}
\def\s{\sigma}
\def\z{\zeta}
\def\om{\omega}
\def\th{\theta}
\def\vth{\vartheta}
\def\vphi{\varphi}

\def\A{{\cal A}}
\def\C{{\bf C}}
\def\D{{\cal D}}
\def\E{{\cal E}}
\def\F{{\cal F}}
\def\G{{\cal G}}
\def\H{{\bf H}}
\def\Hb{{\bf H}}
\def\h{{\cal H}}
\def\I{{\cal I}}
\def\J{{\bf J}}
\def\Jb{{\bf J}}
\def\K{{\cal K}}
\def\Kb{{\bf K}}
\def\L{{\cal L}}
\def\M{{\cal M}}
\def\N{{\cal N}}
\def\Mb{{\bf M}}
\def\Ob{{\bf O}}
\def\O{{\cal O}}
\def\P{{\cal P}}
\def\Q{{\cal Q}}
\def\R{{\bf R}}
\def\S{{\cal S}}
\def\T{{\rm T}}
\def\V{{\cal V}}
\def\W{{\cal W}}
\def\X{{\cal X}}
\def\Z{{\bf Z}}
\def\toro{{\bf T}}

\def\Ga{\Gamma}
\def\De{\Delta}
\def\La{\Lambda}
\def\Om{\Omega}
\def\Th{\Theta}

\def\pa{\partial}
\def\pd{\partial}
\def\d{{\rm d}}       
\def\w{\wedge}
\def\xb{{\bf x}}
\def\x{\times}
\def\ox{\otimes}
\def\o+{\oplus}
\def\xd{{\dot x}}
\def\yd{{\dot y}}
\def\grad{\nabla}     
\def\lapl{\triangle}  
\def\ss{\subset}
\def\sse{\subseteq}
\def\Ker{{\rm Ker}}
\def\Ran{{\rm Ran}}
\def\ker{{\rm Ker}}
\def\ran{{\rm Ran}}
\def\unm{{1 \over 2}}
\def\iff{{\rm iff\ }}
\def\all{\forall}
\def\LRA{\Leftrightarrow}
\def\<{\langle}
\def\>{\rangle}
\def\eor{{$\odot$}}
\def\EOR{~ \hfill {$\odot$}}
\def\EOP{~ \hfill {$\diamondsuit$} \medskip }

\def\shcomp{\odot}
\def\pe#1#2#3{{\d x^{#1} \w \d x^{#2} \w \d x^{#3} }}

\def\interno{\hskip 2pt \vbox{\hbox{\vbox to .18
truecm{\vfill\hbox to .25 truecm
{\hfill\hfill}\vfill}\vrule}\hrule}\hskip 2 pt}

\def\({\left(}
\def\){\right)}
\def\[{\left[}
\def\]{\right]}
\def\=#1{\bar #1}
\def\~#1{\widetilde #1}
\def\.#1{\dot #1}
\def\^#1{\widehat #1}
\def\"#1{\ddot #1}

\def\mapright#1{\smash{\mathop{\longrightarrow}\limits^{#1}}}
\def\mapdown#1{\Big\downarrow\rlap{$\vcenter{\hbox{$\scriptstyle#1$}}$}}
\def\mapleft#1{\smash{\mathop{\longleftarrow}\limits^{#1}}}
\def\mapup#1{\Big\uparrow\rlap{$\vcenter{\hbox{$\scriptstyle#1$}}$}}

\def\mapse#1{\smash{\mathop{\searrow}\limits^{#1}}}
\def\mapnw#1{\smash{\mathop{\nwarrow}\limits^{#1}}}
\def\mapsw#1{\smash{\mathop{\swarrow}\limits^{#1}}}
\def\mapne#1{\smash{\mathop{\nearrow}\limits^{#1}}}

\def\ket#1{{| {#1} \rangle}}

\def\en#1{\eqno(#1)}
\def\ref#1{\cite{#1}}
\def\Proof{{\bf Proof.}~~}
\def\Remark#1{\medskip \noindent {\bf Remark {#1}}}


\thispagestyle{empty}

\FirstPageHead{**}{**}{2005}{\pageref{gaeta-firstpage}--\pageref{gaeta-lastpage}}{Article}

\copyrightnote{2005}{G~Gaeta,~R~Mancinelli}

\Name{Asymptotic scaling in a model class of anomalous
reaction-diffusion equations} \label{gaeta-firstpage}

\Author{Giuseppe GAETA}
\Address{Dipartimento di Matematica,
Universit\`a di Milano, v. Saldini 50, I--20133 Milano (Italy);
{\tt gaeta@mat.unimi.it}}

\Author{Rosaria MANCINELLI}
\Address{Dipartimento di Fisica,
Universit\`a di Roma Tre, Via della Vasca Navale 84, I--00146 Roma
(Italy); {\tt mancinelli@fis.uniroma3.it}}

\Date{Received: \giorno}

\begin{abstract}
\noindent We analyze asymptotic scaling properties of a model
class of anomalous reaction-diffusion (ARD) equations. Numerical
experiments show that solutions to these have, for large $t$, well
defined scaling properties. We suggest a general framework to
analyze asymptotic symmetry properties; this provides an
analytical explanation of the observed asymptotic scaling
properties for the considered ARD equations.
\end{abstract}

\section*{Introduction}

In this note we will consider a class \cite{MVV2} of scalar
partial differential equations (PDEs) of reaction-diffusion type
associated to anomalous diffusion (see e.g. \cite{CLV} for a
recent review focusing on aspects of interest here). This is far
from representing the most general anomalous reaction-diffusion
(ARD) type of equation, but display a variety of behaviors common
to much more general ARD equations.

Numerical experiments on representatives of this class \cite{MVV2}
show that for large $t$ solutions are described by travelling
fronts with a well defined scaling behavior (see below for
details); our goal is to provide an analytical explanation for
this.

In order to do this, we will first recall standard notions from
symmetry analysis of differential equations (sect.1), and then
extend them to the asymptotic framework (sect.2). We will then be
able to propose a general approach to extract asymptotic behavior
of equations based on maps to equations with known asymptotic
symmetry properties (sect.3); the basic idea will be, in the
renormalization group langauge, to identify an equation in the
same universality class amenable to asymptotic analysis.

This approach will be used to analyze the asymptotic behavior of
our class of anomalous reaction-diffusion equations in terms of
the known asymptotic behavior of the FKPP equation. Our results
provide a sound theoretical explanation of the behavior observed
in numerical experiments \cite{MVV2} and recalled below.

\bigskip\noindent
{\bf Acknowledgements.} We thank D. Levi for useful discussions.
The work of GG was supported in part by {\it GNFM--INdAM} under
the program {``Simmetrie e tecniche di riduzione''}; the work of
RM was supported by {\it INFM. -- Istituto Nazionale di Fisica
della Materia}.

\section{Symmetries of differential equations}

We assume the reader to be familiar with the main concepts and
definitions for symmetries of differential equations (see e.g.
\cite{Gae,Olv1,Ste,Win} for general treatments of these), to be
later extended to asymptotic symmetries. In this section we will
fix some general notation to be freely used later on, and recall
general results we need later on; as in this note we will only
consider scalar PDEs, we will specialize formulas to this case.

\subsubsection*{General notation}

We consider an evolution PDE of order $n$ for a real dependent
variable $u = u (x,t)$, with $x,t$ independent real variables. We
denote by $M = \R^2 \times \R$ the total space of independent and
dependent variables, and by $J^{(n)} M$ the jet space of order $n$
over $M$. Given a function $u = f(x,t)$, we denote its graph as
$\ga_f \ss M$, and its prolongation as $\ga_f^{(n)} \ss J^{(n)}
M$.

Let us consider a vector field
$$ X \ = \ \xi (x,t,u) \, \frac{\pa}{\pa x} \ + \ \tau (x,t,u) \frac{\pa}{\pa t} \ + \ \phi (x,t,u) \,  \frac{\pa}{\pa u}   \eqno(1.1) $$
in $M$. For $\eps$ sufficiently small, the function $u = f(x,t)$
is mapped by $\exp (\eps X)$ into a new function $u = \~f (x,t)$;
one obtains with standard computations that this new function is
given by
$$ \~f \ = \ f \ + \ \eps \[ \, \phi - \({\pa f \over \pa x} \) \cdot  \xi - \({\pa f \over \pa t} \) \cdot  \tau \, \]_{u=f(x,t)} \ + \ o (\eps) \ . \eqno(1.2) $$

The action of $X$ in $J^{(n)} M$ is described by its prolongation
$X^{(n)}$. We say that $X$ is a {\it symmetry} of a given equation
if it maps any solution into a (generally, different) solution
\cite{GaeK,Olv1,Ste,Win}. If the equation is given by $\Phi=0$,
this is equivalent to $ [ X^{(n)} (\Phi) ]_{\Phi=0} =  0$.

Geometrically, $\De$ identifies a {\it solution manifold} $S$ in
$J^{(n)} M$; the  function $u = f(x,t)$ is a solution to $\De$ if
and only if $\ga_f^{(n)} \ss S$, and $X$ is a symmetry of $\De$ if
and only if $X^{(n)} : S \to \T S$.

\subsubsection*{Invariant solutions}

If $\De$, which we write as $\Phi (x,t,u,...) = 0 $, is a PDE for
$u = u(x,t)$ admitting a vector field $X$ as symmetry, there is a
well known procedure to determine $X$-invariant solutions to
$\De$; this represents an extension of the familiar method of
characteristics to solve quasilinear PDEs (for more details, see
e.g. the discussion in chap.3 of \cite{Olv1}).\footnote{This
method requires a transversality condition, generically satisfied
for scalar equations as those we wish to consider; see
\cite{AFT,GTW} for discussion and a more general approach.}

First of all we pass to symmetry-adapted coordinates $(y,v,\s)$ in
$M$; the $(y,v)$ will be $X$-invariant coordinates, while $\s$
will be acted upon by $X$. We see $(y,\s)$ as independent
variables and $v$ as the dependent one; we can then use the chain
rule to express $x$ and $t$ derivatives of $u$ in terms of the
$\s$ and $y$ derivatives of $v$, and write $\De$ in terms of the
$(\s,y,v)$ coordinates, i.e. in the form $\^\Phi (\s , y , v,...
)=0$. As $X$ is a symmetry of $\De$, it follows that the function
$\^\Phi$, when subject to the side condition $\pa v / \pa \s = 0$,
is independent of $\s$.

If we are able to determine a solution $v = \^f (y)$ to the
reduced equation, by writing this in terms of the $(x,t,u)$
coordinates we get a $X$-invariant solution to $\De$.

\subsubsection*{Equivalent equations}

Let us consider a map $\chi : M \to M$ which is not a symmetry of
$\De := \De_0$; we write $\chi : (x,t,u) \mapsto (y,s,w)$ and
$\chi^{(n)} : \De \mapsto \^\De$.

If $\De$ is of the form $\De \equiv u_t - F (x,t,u,u_x,u_{xx} ) =
0$ and $\chi$ is projectable \cite{GaeK}, i.e. such that $s = s
(t)$, $y = y(x,t)$, then $\^\De$ is of the same type: $\^\De
\equiv w_s - G (y,s,w,w_y,w_{y y} ) = 0$. Moreover, if $\chi$ is
projectable it maps $\ga_f \ss M$ into a manifold which also is
the graph of a function $w = g (y,s)$, say $\chi (\ga_f) = \ga_g
\ss M$; this extends to prolongations, i.e. $\chi^{(2)} :
\ga_f^{(2)} \to \ga_g^{(2)}$. Hence solutions $u = f (x,t)$ to
$\De$ are mapped into solutions $w = g (y , s)$ to $\^\De$. We say
therefore that $\chi$ is a {\it solution preserving} map
\cite{Wal}.

If $\chi$ is invertible (with a ${\cal C}^n$ inverse), we can
repeat these considerations for $\chi^{-1}$. In this case the
equations $\De$  and $\^\De$ of order $n$ are {\it equivalent}, in
that there is a ${\cal C}^n$ isomorphism between solutions to
$\De$ and solutions to $\^\De$.

\subsubsection*{Conditional and partial symmetries}

As mentioned above, $X$ is a symmetry if it maps any solution to a
(generally, different) solution. However, there can be cases where
this is true only for some solution; one can formulate
correspondingly weaker notions of symmetry. (For a review of
different ``extended'' notions of symmetry, see \cite{OlV}; see
also \cite{CiK} for a recent and shorter discussion.)

We say that $X$ is a {\it partial symmetry} for $\De$ if there is
a nonempty set ${\cal S}_X$ of solutions to $\De$ which is mapped
to itself by $X$. The set ${\cal S}_X$ could be made of a single
solution, and more generally that there could be solutions which
are left invariant by $X$. In this latter case, we say that $X$ is
a {\it conditional symmetry} for $\De$.

It follows from (1.2) that $u = f(x,t)$ is invariant under $X$ if and only if
$$ \De_X \ := \ \phi [x,t,f(x,t)] - u_x \xi [x,t,f(x,t)] - u_t \tau [x,t,f(x,t)] \ = \ 0 \ ; \eqno(1.3) $$
thus $X$-invariant solutions to $\De$ are solutions to the system
made of $\De$ and $\De_X$. In other words, $X$ is a conditional
symmetry of $\De$ if and only if it is a symmetry of this system
(and is not a proper symmetry of $\De$).

Partial symmetries can be seen in a similar way. As discussed in
\cite{CGpar}, a function $f$ in the globally invariant set of
solutions to $\De : \Phi = 0$  ($ f \in {\cal S}_X$ in the
notation used above) will be a solution to a system
$$ \Phi^{(0)} = 0 \ , \  \Phi^{(1)} = 0 \ , \  ...... \ , \ \Phi^{(p)}
= 0 \ ,  $$ where $\Phi^{(0)} \equiv \Phi$, and $\Phi^{(k+1)} := Y
[ \Phi^{(k)} ]$. The integer $p$, i.e. the order of the system, is
determined as the lowest order such that $\Phi^{(p)}$ vanishes
identically on common solutions to all the previous equations.
Each equation $\Phi^{(k)}=0$ can, and should, be simplified by
taking into account the previous equations; for concrete examples,
see \cite{CGpar}.

If $\De$ and $\^\De$ are equivalent via a solution-preserving map
$\chi$, this entails a corresponding relation between their
conditional and/or partial (as well as ordinary) symmetries.

\section{Asymptotic symmetries of PDEs}

The concept of symmetry of a differential equation can be extended
to that of asymptotic symmetry; we will again confine ourselves to
scalar second order evolution PDEs. Asymptotic symmetries (in the
present sense) were introduced and discussed in quite a general
framework in \cite{Gae}, see also \cite{CG}.

As suggested by their names, asymptotic symmetries of a
differential equation $\De$ are vector fields $X$ which, albeit in
general not symmetries of $\De$, satisfy $X^{(2)} : S_\De \to \T
S_\De$ asymptotically (see below for the precise sense of this).
Any exact symmetry is also (trivially) an asymptotic symmetry.

We write the equation $\De$ in the form
$$\Phi \equiv u_t - F (x,t,u,u_x,u_{xx}) = 0 \ , \eqno(2.1) $$ and denote by $\F$ the space of maps $F : (x,t,u,u_x,u_{xx}) \to \R$ which are polynomial in $(u,u_x,u_{xx})$ and rational in $(x,t)$. Note that this space, which corresponds to the space of equations in the form we are considering, is invariant under scaling transformations and under translations.

Vector fields will be written as in (1.1). The second prolongation
of $X$ will be denoted as $Y \equiv X^{(2)}$; as seen above, if
$X$ is projectable, then $Y$ acts in $\F$ \cite{GaeK}.

We write the flow (in the space of functions $f:\R^2 \to \R$)
issued by $f_0 (x,t)$ as $ f_\la (x,t) = \exp [{\la \^X}]$, with
$d f_\la / d \la = \^X [ f_\la ] = [\phi - f_x \xi - f_t
\tau]_{u=f(x,t)}$, see (1.2). We say that $f_0 (x,t)$ is
$X$-invariant if it is a fixed point for the flow of $\^X$, i.e.
if $\^X [ f_0 ] = 0$.

It may happen that $f_0$ is not $X$-invariant, but the flow
$f_\la$ is asymptotic to an invariant function $f_*$, i.e.
$$ \lim_{\la \to \infty} | f_\la (x,t) - f_* (x,t) | = 0 \ \ , \ \ \^X [ f_* ] = 0 \ . \eqno(2.2) $$
When (2.2) is satisfied, we say that $f$ is {\it asymptotically
$X$-invariant} under the flow of $\^X$.

Let us now consider the action of $X$, more precisely of $Y$, on
the space of equations in the form (2.1), i.e. in $\F$. We write
$$ \De_\la \ = \ e^{\la Y} \De_0 \ := \ \s (\la) \, \[ u_t - F_\la (x,u,u_x,u_{xx} ) \]   \eqno(2.3) $$
for the $Y$ flow issued from $\De = \De_0$; by construction, this
satisfies $ d \De_\la / d \la  = Y ( \De_\la )$. In this way, $X$
induces a vector field $W$ in the space $\F$ ($W$ is nothing else
than the restriction of $Y$ to $\F$); thus (2.3) is equivalent to
$d F / d \la = W (F)$.

As recalled above, $X$ is a symmetry of $\De$ if and only if $Y :
S \to \T S$. This condition is now rephrased in terms of $\F$ by
saying that $X$ is a symmetry of $\De_0$ given by $\De_0 := u_t -
F_0 (x,t,u,...)$ if and only if $F_0$ is a fixed point for the
flow of $W$.

Suppose now that $X$ is not a symmetry of $\De_0$ -- i.e. $F_0$ is not a fixed point for the flow of $W$ in $\F$ -- but that the flow $F_\la$ issued by $F_0$ under $e^{\la W}$ satisfies
$$ \lim_{\la\to \infty} | F_\la - F_* | \ = \ 0 \ \ , \ W (F_* ) = 0 \ . $$
In this case we say that $X$ is an {\it asymptotic symmetry} for
$F_0$, i.e. for the equation $\De_0$.

By construction and by the invertibility of the map $\exp (\la X)$ -- hence of $\exp (\la Y)$ -- for $\la$ finite, if $u = f (x,t)$ is a solution to $\De_0$ then $u = f_\la (x,t)$ will be a solution to $\De_\la$; and conversely if $u = f_\la (x,t)$ is a solution to $\De_\la$ then $u = f (x,t)$ is a solution to $\De_0$.

In the limit $\la \to \infty$ the invertibility of (2.3) fails if
the flow goes to a limit point; we can nevertheless still say that
solutions to the original equation flow into solutions to the
asymptotic equation, provided both limits $f_*$ and $\De_*$ exist.

If the limit $\De_*$ exists, it captures the behavior of $\De$ for
large $\la$, i.e. for large (or small, depending on the sign of
$a$ and $b$) $t$ and $|x|$.

\subsection*{Solution-preserving maps and asymptotic symmetries}

The use of this construction, combining with solution-preserving
maps, is the following.

If there is an equation $\^\De$ whose asymptotic behavior is well
understood, and such that there exists a solution-preserving map
$\chi$ with $\chi^{(2)} : \De \to \^\De$, we can study the
asymptotic behavior of solutions $u(x,t)$ to $\De$ by means of the
asymptotic behavior of solutions $w (y,s)$ to $\^\De$.

If $\chi$ has a smooth inverse, and $W$ is mapped by $\chi^{(2)}$
into $\^W$, we can study the flow of $W$ using $$ e^{\la W} \ = \
[ \chi^{(2)} ]_{\F}^{-1} \, \circ \, e^{\la \^W} \, \circ \,
\chi^{(2)} \ . \eqno(2.4) $$ Note that (2.4) remains valid also in
the limit $\la \to \infty$, i.e. for the asymptotic regime.

Needless to say, this approach is particularly convenient when the
asymptotic behavior of the solutions $w (y,s) $ to $\^\De_0$ --
i.e. the behavior of solutions to $\^\De_*$ -- is simple and/or in
some way universal (even better if $\^\De_0$ is a fixed point
under $\^W$).

Denoting as $w_* (y,s)$ the limit expression for the solutions to
$\^\De_0$, the asymptotic behavior of the solution $u (x,t)$ to
$\De_0$ will be given by
$$ u_* (x,t) \ = \ \chi^{-1} \, \[ w_* (y,s) \] \ . \eqno(2.5) $$
More precisely, considering $\ga_f$ and $\ga_g$ corresponding to
$u = f (x,t)$ and $w = g (y,s)$ (see sect.1), and going
asymptotically into $\ga_f^*$ and $\ga_g^*$ respectively, we have
$$ \ga_f^* \ = \ \chi^{-1} \, \( \ga_g^* \) \ . \eqno(2.5') $$

\subsubsection*{Conditional and partial asymptotic symmetries}

It is also possible to consider asymptotic versions of conditional
and partial symmetries.

Let the function $u = f_0 (x,t)$ (more precisely, the
corresponding graph $\ga_0 \ss M$) flow under $e^{\la X}$ to a
fixed point $u = f_* (x,t)$ (more precisely, a $X$-invariant graph
$\ga_* \ss M$) albeit $\De_0$ does {\it not} flow to a fixed
point.

In such a situation, the solution manifold $S_\la \ss J^{(2)} M$
does not go to a limit manifold, but there is a submanifold
$S_\la^X \ss S_\la$, with $\ga_\la \sse S_\la^X \ss S_\la$, which
flows to a fixed limit submanifold $S_*^X$, with $\ga_* \sse
S_*^X$. In this case we say that $X$ is a {\bf conditional
asymptotic symmetry} for $\De$.

The same construction, with suitable and rather obvious modifications, applies for what concerns partial symmetries. Suppose that $\De \equiv \De^{(0)}$ does not flow to a fixed point under $W$, and consider the equations
$$ \De^{(1)} := Y [ \De^{(0)} ] \ , \ ... \ , \ \De^{(r)} := Y [ \De^{(r-1)} ] $$
up to an $r$ -- if it exists -- such that $\De^{(r)}$ does admit a fixed point $\De^{(r)}_*$ under the $W$ flow, while the $\De^{(k)}$ with $k < r$ do not. Then the manifold
$$ S_\la^{(0)} \cap ... \cap S_\la^{(r)} \ := \ {\cal S}_\la $$
(with $S_\la^{(k)} \ss J^{(2)} M$ the solution manifold for $\De_\la^{(k)}$) flows to a limit submanifold ${\cal S}_*$, and solutions $u = f_0 (x,t)$ to the system $\De^{(k)}$ ($k=0,...,r$) flow to functions $u = f_* (x,t)$ such that the prolongation $\ga_*^{(2)} \ss J^{(2)} M$ of the corresponding graph $\ga_{f_*} = \{ (x,t,f_* (x,t)) \}$ lie in ${\cal S}_*$.
In this case we say that $X$ is a {\bf partial asymptotic symmetry} for $\De$.

If $\De$ and $\^\De$ are related by a solution-preserving map
$\chi^{(2)}$, then $\chi$ also relates their conditional and
partial symmetries. In particular, if $\chi$ is invertible and
$\^\De$ admits $X$ as a conditional symmetry, then $\De$ admits
$\chi^{-1}_* (X)$ as a conditional symmetry, with $\chi^{-1}_*$
the extension of $\chi^{-1}$ to vector fields. This will be of use
below.

\section{Asymptotic symmetries as a tool to test asymptotic \\ behavior}

In many physically relevant cases, one has to study nonlinear PDEs
which are not amenable to an exact treatment, or at least for
which such a treatment is not known, and for which a numerical
study shows an asymptotic behavior which appears to be well
described by some kind of invariance. Usually, the latter
corresponds to a scale invariance (self-similar solutions), or a
translation invariance (travelling waves), or a combination of
both of these.

The discussion conducted so far can be implemented into a
procedure allowing on the one hand to test if the observed
asymptotic behavior is a characteristic of the equation (rather
than an artifact of the numerical experiments conducted on it),
and on the other hand to formulate simpler equations extracting
the asymptotic behavior.

We will now describe the procedure in operational terms; as a
(rather simple, but relevant) example to illustrate our procedure
we will consider here the heat equation, and then anomalous
diffusion equations, while in later sections we apply the
procedure on anomalous reaction-diffusion equations.

We denote by $X$ the vector field describing the observed
invariance, and consider a second order  equation for $u = u(x,t)$
of the form $\De :=  u_t - F (x,t,u,u_x,u_{xx}) = 0$ (hence $M =
\{ (x,t;u) \}$).

\begin{itemize}

\item {\bf Step 1.} Pass to symmetry-adapted coordinates in $M$,
i.e. coordinates $(\s , y ; v)$ such that $X(y) = X(v) = 0$; thus
in these coordinates $X = f(\s,y,v) (\pa / \pa \s )$. \footnote{We
stress that we are not requiring $f=1$; actually when we deal with
scaling symmetries it is appropriate to require $f (\s , y , v) =
\s$.}

\item {\bf Step 2.} Identify $v$ as the new dependent variable,
i.e. $v = v (\s , y)$. This allows to write $x$ and $t$
derivatives of $u$ as $\s$ and $y$ derivatives of $v$, hence to
write the differential equation $\De (x , t;u^{(2)} ) $ as $\^\De
(\s,y;v^{(2)} )$.

\item {\bf Step 3.} Reduce the equation $\^\De (\s,y;v^{(2)} )$ to
the space of $X$-invariant functions, i.e. to $v(\s,y)$ satisfying
$v_\s = 0$. For $X$ an exact symmetry, the reduced equation
$\^\De_X$ will not depend on $\s$ at all. For $X$ a conditional or
partial symmetry, $\s$ will still appear parametrically in the
reduced equation.

\item {\bf Step 4.} Study the asymptotic behavior of the solutions
to the reduced equation $\^\De_X$ for $\s \to \infty$.

\item {\bf Step 5.} Go back to the original variables.

\end{itemize}

\subsubsection*{Elementary example: the heat equation}

Let us illustrate our procedure by applying it on the heat equation $u_t = u_{xx}$. Its asymptotic solutions are of the form
$$ u(x,t) \ = \ {A \over \sqrt{2 t} } \ \exp \[ - {4 x^2 \over t} \] \eqno(3.1) $$
and are invariant under the scaling vector field
$$ X \ = \ x \pa_x \, + \, 2 t \pa_t \, - \, u \pa_u \ , \eqno(3.2) $$
which is also an exact symmetry of the equation.

The symmetry-adapted coordinates are $\s = t$, $y = x^2 /t$, and
$v = x u$. In these, the heat equation reads
$$ \s \, v_\s \ = \ 4 \, y \, v_{yy} \ + \ (2 + y) \, v_y \ + \ v \ . \eqno(3.3) $$
Note that (3.3) necessarily requires that for $\s \to \infty$,
$v_\s = 0$: this shows in a simple way that the full asymptotic
behavior of (3.3) is captured by (3.1).

Imposing $v_\s = 0$ in (3.3), $\s$ disappears completely. Needless
to say, the equation obtained in this way has solutions $ v (y) =
\widehat{A} \sqrt{y} \exp [- y/4]$, which when mapped back to the
original coordinates produce the gaussian (3.1).

\subsubsection*{Example: anomalous diffusion equations}

The procedure described above can also be applied to what will be
our model class of anomalous diffusion equations, studied
numerically in \cite{MVV2}; these are written as
$$ u_t \ = \ {x^{1 - \a/2} \over t^{1 - \nu \a} } \ {\pa \over \pa x} \[ x^{1 - \a/2} u_x \] \ . \eqno(3.4) $$
To focus ideas, let us mention two examples of equations in this
class: {\bf (i)} For $\a = 2$ we have the generalized gaussian
process; {\bf (ii)} For $\a = 2/3$, $\nu = 3/2$  we deal with the
Richardson equation describing the evolution of the distance
between two particles in developed turbulent regime; we refer to
\cite{MVV2} for a discussion of the interest of the class of
anomalous RD equations (3.4).

One can check that the map
$$ s = t^{\a \nu} \ , \ y = x^{\a/2} \ , \ w = t^{(2-\a)(\nu/2)} u  \eqno(3.5) $$
is solution preserving from (3.4) to the heat equation $ w_s \ = \
w_{y y} $. Using the inverse change of coordinates, the universal
asymptotic solution $ w (y,s ) \simeq s^{- 1/2} e^{-y^2/s}$ of the
heat equation is mapped back into
$$ u(x,t) \ \simeq \ {1 \over t^\nu} \ \exp \[ - x^\a \, / \, t^\nu \] \ , \eqno(3.6) $$
which represents therefore the universal asymptotic solution to
(3.4). This result is confirmed by numerical experiments
\cite{MVV2}.

\section{The FKPP equation: asymptotic solutions, and \\ symmetries}

In the same way as in the previous example the heat equation
played the role of target for solution preserving maps applied to
anomalous diffusion equations, in the case of (our model class of)
anomalous reaction-diffusion (ARD) equations we will look for a
solution-preserving map to the well known
Fisher-Kolmogorov-Petrovskii-Piskunov (FKPP) equation
\cite{Fis,KPP,Mur}. This reads
$$ u_t \ = \ D u_{xx} \, + \, \eps u (1-u) \ ,  $$
with $\eps$ and $D$ real positive constants, and one requires that
$u(x,t) \ge 0$ for all $x$ and $t$. There are two stationary
homogeneous states, i.e. $u = 0$ and $u=1$; the latter is stable
while the former is unstable against small perturbations. In this
section we study asymptotic symmetries of the equation and of its
solutions.

It is well known \cite{KPP} that if the initial datum is suitably
concentrated, e.g. $u (x,0) = 0$ for $|x-x_0| > L$ or more
generally $u(x,0) < A \exp [ - x / L ]$, then asymptotically for
$t \to \infty$ and $x \to \infty$ the solution is of the form $ u
= f(x,t) \simeq \exp [ - (x - v t) / \la ]$, with $\la =
\sqrt{D/\eps}$ and $v = \sqrt{4 \eps D}$. This  represents a {\it
front} of width $\la$ travelling with speed $v$; it connects the
stable state $u=1$ and the unstable state $u=0$.

In discussing the FKPP equation, it is convenient to pass to
rescaled coordinates $ \~t = \eps t$, $\~x = (\sqrt{\eps / D}) x$.
From now on we will use these coordinates, and omit the tildas for
ease of notation. In these coordinates, the FKPP equation reads
$$ u_t \ = \ u_{xx} + u (1-u) \ . \eqno(4.1) $$
As for the asymptotic solution described above, this reads now
$$u = f_0 (x,t) \ \simeq \ A \ \exp \[ - (x - 2 t) \] \ ; \eqno(4.2) $$
note the front has speed $v=2$ and width $\la=1$.

It should be stressed that the $f(x,t)$ or $f_0 (x,t)$ given above
provide the solution for $x \to \infty$, i.e. in the region of
small $u$; in this region (4.1) is well approximated by its
linearization around $u=0$, i.e.
$$ u_t = u_{xx} + u \ ; \eqno(4.3) $$
the ansatz $u (x,t) = w(z) := w (x - 2 t)$ takes this into the ODE
$$ w'' + 2 w' + w \ = \ 0  \eqno(4.4) $$
for $w = w(z)$, with solution
$$ w (z) \ = \ c_1 e^{-z} + c_2 z e^{-z} \ . \eqno(4.5) $$
We denote by ${\cal W}$ the set of solutions described by (4.5);
note that ${\cal W} = \R^2$, and $(c_1,c_2)$ provide coordinates
in ${\cal W}$.

The $f_0$ given above, see (4.2), corresponds to $c_2 = 0$. This
can be characterized in terms of symmetry properties, as discussed
below. It is convenient to consider linear combinations of the
shifts, given by $X_\pm = X_1 \mp (1/2) X_2$; note that $z = x - 2
t$ is invariant under $X_-$, and that $X_+ = \pa_z$. We also have
$X_0 = w \pa_w$. Needless to say, $[X_0,X_+] = 0$.

\medskip\noindent
{\bf Lemma 1.} {\it The symmetry algebra of the linearized
equation (4.3) is generated by the scaling $X_0 = u \pa_u$, and by
the translations $X_1 = \pa_x$ and $X_2 = \pa_t$. The quotient
equation (4.4) admits only $X_0$ as scaling symmetry; it also
admits the translation symmetry generated by $X_+$, while $X_-$
has been quotiented out by passing to the $z$ variable.}

\medskip\noindent
{\bf Proof.} This follows from standard (and elementary) computations. \EOP

\medskip\noindent
{\bf Lemma 2.} {\it The propagating front solutions correspond to
a subspace of ${\cal W}$ invariant under the action of the group
generated by the vector fields $X_0$ and $X_+$.}

\medskip\noindent
{\bf Proof.} A general element of the group generated by $X_0$ and
$X_+$ is written as $g (\a , \b) := \exp [ \a X_0 + \b X_+ ]$ and
acts on ${\cal W}$ by
$$ g (\a,\b) \ : \ (c_1,c_2) \ \to \ \(  e^{\a + \b} (c_1 + \b c_2 ) \, , \,
e^{\a + \b} c_2 \) \ . $$ The subspace $c_2 = 0$ is invariant
under this action. As remarked above, the propagating front
solutions (7.2) correspond to $c_2 = 0$. \EOP

It is immediate to see that the only scaling or shift symmetries
of the full FKPP equation (4.1) are those, with generators $X_1 =
\pa_x$ and $X_2 = \pa_t$, corresponding to translations in $x$ and
$t$; these reflect the fact that (4.1) is a homogeneous equation.

The situation is different for what concerns asymptotic
symmetries, and in particular scaling ones, as we now discuss.

\medskip\noindent
{\bf Lemma 3.} {\it Let $X$ be a scaling vector field, such that
$\lim_{\la \to \infty} \exp (\la X)$ extracts the behavior for
large $|x|$ and $t$. Let $\De_0$ be the FKPP equation, and
$\De_\la = e^{\la Y} \De_0$ with $Y$ the prolongation of $X$. Then
$\lim_{\la \to \infty} \De_\la = \De_*$ is the heat equation $u_t
- u_{xx}$.}

\medskip\noindent
{\bf Proof.} We consider the most general scaling generator, i.e.
a vector field in the form (1.3), $ X = a x \pa_x + b t \pa_t + c
u \pa_u$. We can always set one of the constants $(a,b,c)$ equal
to unity (provided it is nonzero); this amounts to a redefinition
of the scaling group parameter.

Applying the procedure described in previous sections, with of
course $\De_0 := u_t - u_{xx} - u (1 -u) = 0$ the FKPP equation,
we obtain at once that
$$ \De_\la \ = \ \la^{c-b} \ \[ u_t \, - \, \la^{b - 2 a} u_{xx} \, - \, \la^b u \, + \, \la^{b+c} u^2 \] \ . \eqno(4.6) $$
We choose $c=b$ and  $a=b/2$. In order for $\lim_{\la \to \infty}
\exp (\la X)$ to extract the behavior for large $|x|$ and $t$, we
must choose $a<0$, $b<0$. We can set the modulus of one of the
constants, say $b$ for definiteness, equal to unity; i.e. $b =
-1$. With these choices, we have
$$ \De_\la \ = \ u_t \, - \, u_{xx} \, - \, \la^{-1} u \, + \, \la^{-2} u^2 \ . \eqno(4.7) $$
The limit $ \De_*  :=  \lim_{\la \to \infty} \De_\la$ is the heat equation $u_t - u_{xx} = 0$, as claimed. \EOP

\section{Anomalous reaction-diffusion equations}

The FKPP is a (relevant) representative of a more general class of
anomalous reaction-diffusion equations, which we write as
$$ u_t \ = \ \^L [u] \, + \, h (u) \ ; \eqno(5.1) $$
here $\^L$ is the linear operator describing passive transport of
the field $u$, hence anomalous diffusion, while $h (u)$ describes
its growth. Logistic growth, which we will consider, corresponds
to choosing $ h(u) := u (1 - u)$. With this choice and considering
the $\^L$ associated to anomalous diffusion in our model class,
see the r.h.s. of (3.4), we get the equation
$$ u_t \ = \ {x^{2-\a } \over t^{1 - \nu \a} } \ \[ u_{xx} \, + \, {(2 - \a ) \over 2 x} \, u_x \] \ + \ u \, (1 - u) \ . \eqno(5.2) $$
One is usually interested in solutions with initial data $u(x,0)$
which are suitably regular and with compact support. A detailed
numerical study of systems of the form (5.2) with such initial
data was conducted in \cite{MVV2}.

We summarize the findings of these numerical experiments as
follows:

\begin{itemize}
\item{{\tt (i)}} asymptotically for large $x$ and $t$, the solution is
described by a travelling front with varying speed $c (t)$ and
width $\la (t)$; the form of this front for small $u$ is well
described by
$$ u (x,t) \ \simeq \ A \ \exp \[ - { x - c(t) \cdot t \over \la (t) } \] \ ; \eqno(5.3) $$

\item{{\tt (ii)}} the (asymptotic) time dependence of $c(t)$ and
$\la (t)$ are described by
$$ c(t) \ \simeq \ c_0 \cdot t^{\de} \ , \ \la (t) \ \simeq \ \la_0 \cdot t^{\de} \eqno(5.4) $$
where $c_0$ and $\la_0$ are dimensional constants;
\item{{\tt (iii)}}
the scaling exponent $\de$ is given by
$$ \de \ := \ \nu \, + \, (1/\a) \, - \, 1 \ . \eqno(5.5) $$
\end{itemize}

\noindent Thus we rewrite (5.3) in the form
$$ u(x,t) \ \simeq \ A \ \exp \[ - { x - (c_0 t^\de) t  \over \la_0 t^\de } \] \ .  \eqno(5.6) $$
Note that for $\de = 0$, i.e. for $\nu = 1 - (1/\a)$, the front
travels with constant speed and width, as for the FKPP equation.

We want now to describe precisely the invariance properties of the
observed asymptotic solution (5.6), in particular for what
concerns scaling transformations.

\medskip\noindent
{\bf Lemma 4.} {\it The scaling invariance of (5.6) is described
by the generalized scaling group
$$
x \to \ (\la^\de) \ x \ , \ \ t \to \ \la \ t \ , \ \ u \to \
\[ \exp \( {(\la - 1) \, K \, t } \) \] \ u \ , \eqno(5.7) $$
with $\la$ the group parameter.}

\medskip\noindent
{\bf Proof.} The generator of the one-parameter group described by
(5.7) is
$$ X \ = \ \de \, x \, \pa_x \ + \ t \, \pa_t \ - \ K t \, u \, \pa_u \ . \eqno(5.8) $$
The invariance of (5.6) under this can be easily checked using
(1.2). It can be shown that this is the only scaling type symmetry
of (5.6). \EOP

\medskip\noindent
{\bf Theorem 1.} {\it The vector field (5.8) is not a symmetry of
the equation (5.2), but it is an asymptotic symmetry of the same
equation.}

\medskip\noindent
{\bf Proof.} Let us denote (5.2) by $\De_0$ and (5.8) by $X$;
applying $Y \equiv X^{(2)}$ on $\De_0$, and restricting to the
solution manifold $S_0$ of $\De_0$ (this amounts to substituting
for $w_\s$ according to $\De_0$ itself), we obtain
$$ \begin{array}{rl}
\De_1 := \ \[ Y (\De_0 ) \]_{S_0} \ =& \ [ (1- \a) t^{\a \de} \(x/t\)^{2-\a} ] \, u_{xx} \ + \\
& \ +  \[ (1/2) (\a - 1 ) (\a - 2) t^{\a \de - 1} \( x/t \)^{1 - \a } \] \, u_x \ + \\
& \ - \ u ( 1 + K - u + K u t )  \ . \end{array} $$ This is not
zero, i.e. $X$ is not a symmetry of (5.2).

Going further on with our procedure, we have to compute $\De_2 \
:= \[ Y (\De_1 ) \]_{S_0 \cap S_1}$. It results
$$ \De_2 \ := \
\[ 2 - 4 K t + K^2 t^2 - \a (1 - K t ) \] \, u^2 \ - \
\[ (1+K)(\a - 2 ) \] u \ = \ 0 \ . $$ This has
the trivial solution $u = 0$, which is also solution to $\De_0$
and $\De_1$, and the nontrivial solution
$$ u (t) \ = \ { (2 - \a ) (1 + K ) \over (2 - \a) - (4 - \a ) K t + K^2 t^2 } \ .  $$
The latter, as easily checked by explicit computation, is in general {\it not} a solution to $\De_0$ and $\De_1$: inserting this into $\De_0$ and $\De_1$ we have respectively
$$ \widetilde{\De_0} \ = \ \[(\a - 2) K (1 + K)\] \ \({K t^2 + (2 K - 4 +  \a ) t + (2 \a  - 6 ) \over \( K^2 t^2 + (\a - 4 K) t + (2 - \a ) \)^2 } \) \ , $$
$$ \widetilde{\De_1} \ = \ \[ (\a - 2) K (1+K)^2 \] \ \( { K t^2 - 2 t \over \( K^2 t^2 + (\a - 4 K ) t + (2 - \a ) \)^2 } \) \ . $$
For $K \not= 0,-1$, both of these expressions are not zero (unless
$\a = 2$, corresponding to gaussian processes). However, both of
these go to zero (like $1/t^2$), for all $\a$ and $K$, in the
limit $t \to \infty$.  \EOP

Having determined that $X$ is an asymptotic (partial) symmetry for
our equation $\De_0$, we will now apply our general procedure. The
first step consists in passing to symmetry adapted coordinates;
these are
$$ \s = t \ , \ \ y = x / t^\de \ , \ \ w = u e^{K t } \ . \eqno(5.9') $$
In these coordinates, the vector field (5.8) reads simply $ X = \s
\pa_\s$,  and the (obviously $X$-invariant) asymptotic solution
(5.6) is $ w = A \exp [ y / \la_0 ]$. With standard computations
we obtain
$$ \begin{array}{rl}
u_t \ =& \ \[ w_\s - \de (y / \s ) w_y - K w \] \ e^{- K \s } \ , \\
u_x \ =& \ \[ (1 / \s^\de ) \, w_y \] \ e^{- K \s } \ , \\
u_{xx} \ =& \ \[ (1 / \s^\de )^2 \, w_{yy} \] \ e^{- K \s } \ .
\end{array} \eqno(5.9'') $$ Using (5.9') and (5.9''), we
express $\De_0$ in the new coordinates; this results to be
$$ w_\s \, = \, \[ {y^{2 - \a} \over \s^{\chi} } \] \ w_{yy}  +  \[ \( {2 - \a \over 2} \)  \( {y^{1-\a} \over \s^{\mu } } \)  +  \a \({y \over \s} \) \] \ w_y  + (K+1) w - e^{- K \s} w^2 \ , \eqno(5.10) $$
where we have defined $\mu  = \a (\de - \nu + 1/ \a )$ for ease of
writing.

The expression (5.10) holds for the general map (5.9); however we
are specially interested in the choice $\de = (\nu -1 + 1/\a )$,
see (5.5). With this, we have $\mu  = (2 - \a)$, and finally (5.2)
reads
$$ w_\s \ = \ \( { y \over \s} \)^{2 - \a} \, w_{yy} \ + \ \[ \( {2 - \a \over 2 \s } \) \({y \over \s }\)^{1-\a} + \a  { y \over \s } \] \, w_y \ + \ (K+1) \, w \ - \ e^{- K \s } \, w^2 \ . \eqno(5.11) $$
In the limit $\s \to \infty$, the last term disappears (faster
than any power in $\s$), and (5.11) reduce to a linear equation.

\medskip\noindent
{\bf Theorem 2.} {\it The equation (5.11) has no nontrivial
$X$-invariant solutions, but admits nontrivial asymptotically
$X$-invariant solutions.}

\medskip\noindent
{\bf Proof.} The $X$-invariant solutions to (5.11) are obtained by
requiring that $w_\s = 0$; with this the equation reduces to
$$ \( { y \over \s} \)^{2 - \a} \, w_{yy} \ + \ \[ \( {2 - \a \over 2 \s } \) \({y \over \s }\)^{1-\a} + \a  { y \over \s } \] \, w_y \ + \ (K+1) \, w \ - \ e^{- K \s } \, w^2 \ = \ 0 \ . \eqno(5.12) $$
Note that $\s$ appears parametrically here, and (5.12) splits into
the equations corresponding to the vanishing of coefficients of
different powers of $\s$ (this is a general feature of partial or
``weak'' symmetries, see \cite{CiK}). The only common solution to
these is $w = 0$, which proves the first part of the statement.

Let us go back to considering (5.11); in order to study its
asymptotic behavior for $\s \to \infty$, we disregard the term
which is exponentially small for large $\s$. The resulting linear
equation for $w = w(y)$, i.e.
$$ \( { y \over \s} \)^{2 - \a} \, w_{yy} \ + \ \[ \( {2 - \a \over 2 \s } \) \({y \over \s }\)^{1-\a} + \a  { y \over \s } \] \, w_y \ + \ (K+1) \, w \  = \ 0 \ , $$
yields as solution
$$ w(y) \ = \ C_1 \, \K [ 0 , 0 ]  \, + \, \( \sqrt{ 2 (K+1) \s^2 (y/\s)^\a } \) \, C_2 \, \K [1/2 , 1 ] $$
where $C_1 , C_2$ are arbitrary constants, and
$$ \K [ x,y ] \ := \ F_{11} \[ (K+1) \s / \a^2 + x ; 1/2 + y ; - \s (y/\s)^\a \] $$ with $F_{11} \equiv {}_1F_1$ the Kummer confluent hypergeometric function
$$ F_{11} [a;b;z] := \, {}_1F_1 (a;b;z) \ = \ { \Ga (b) \over \Ga (b-a) \, \Ga (a) } \ \int_0^1 \, e^{z t} \, t^{a-1} \, (1-t)^{b-a-1} \, {\rm d} t \ . $$
The asymptotic solution could now be expressed in terms of the
original variables using (5.9); this yields an explicit but
involved and not specially illuminating expression. \EOP

\section{Other asymptotic partial symmetries of ARD equations}

The scaling symmetry (5.8) is not the only symmetry of the
observed asymptotic solution (5.6) to (5.2). In this section we
identify different symmetries and apply our approach on the basis
of these.

\medskip\noindent
{\bf Lemma 5.} {\it The vector field $X = \xi \pa_x + \tau \pa_t +
\phi \pa_u $ is a symmetry of the asymptotic solution (5.6) to the
equation (5.2) if and only if it belongs to the two dimensional
module (over smooth real functions ${\cal C}^\infty (\R^3 , \R)$
of $x,t,u$) generated by
$$ X_1 \ = \ \pa_x + \( {1 \over \la_0 t^\de } \) \pa_u \ ; \
X_2 \ = \ \( \la_0 t^{1 + \de } \) \pa_t - \( x \de + c_0 t^{1 +
\de} \) \pa_u \ . \eqno(6.1) $$}

\medskip\noindent
{\bf Proof.} This follows easily by using (1.2) and the explicit
expression (5.6) of the asymptotic solution $u = f_* (x,t)$.
Indeed, applying (1.2) we get
$$ {(\~f - f ) \over \eps } \ = \ \phi \, - \, A \, \exp \[ (c_0 t - x t^{- \de})/\la_0 \] \, { (x \de + c_0 t^{1 + \de} ) \tau - t \xi \over \la_0 t^{1 + \de } } \ , $$
and the result follows immediately \EOP

In the following, we will consider in particular
$$ X_0 \ := \ \( x \de + c_0 t^{1+\de} \) \pa_x + t \pa_t \ ,  \eqno(6.2) $$
as well as $X_1$ and $X_2$ themselves. Note that $X_0$ operates in
the space of independent variables alone.

We will write second-prolonged vector fields in the form
$$ Y \equiv X^{(2)} \ = \ X \ + \ \Psi_x {\pa \over \pa u_x} + \Psi_t {\pa \over \pa u_t} + \Psi_{xx} {\pa \over \pa u_{xx}} + \Psi_{xt} {\pa \over \pa u_{xt}} + \Psi_{tt} {\pa \over \pa u_{tt}} \ ;  \eqno(6.3) $$
in view of (5.2) we will actually need only the coefficients
$\Psi_x , \Psi_t , \Psi_{xx} $. We will, as in the previous
discussion, denote (5.2) as $\De_0$.

\medskip\noindent
{\bf Theorem 3.} {\it The vector fields $X_0$, $X_1$ and $X_2$ are
partial symmetries of $\De_0$.}

\medskip\noindent
{\bf Proof.} We will denote by $Y_i$ the second prolongations of $X_i$. Let us start by considering $X_1$.
In this case the coefficients of the second-prolonged vector field $Y_1$ are:
$$ \Psi_x = 0 \ , \ \Psi_t = - (\de / \la_0) t^{-(1+\de)} ) \ , \ \Psi_{xx} = 0 \ , \ \Psi_{xy} = 0 , \Psi_{tt} = (\de / \la_0 ) (1 + \de) t^{-(2 + \de )} \ .  $$
We define then $ \De_1 = \[ Y_1 (\De_0) \]_{S_0}$; $\De_2 =
\[ Y_1 (\De_1) \]_{S_0 \cap S_1}$. By explicit computation,
it results that
$$ \[ Y_1 (\De_2) \]_{S_0 \cap S_1 \cap S_2} \ = \ 0 \ ; $$
this shows that $X_1$ is a partial symmetry for $\De_0$.

For the other vector fields, note that the relevant coefficients of $Y_2$ are
$$ \Psi_x = - {\de \over \la_0 t^{1 + \de }} \ , \ \Psi_t = {(1 + \de ) \de x \over \la_0 t^{2 + \de} } \ , \ \Psi_{xx} = 0 \ , $$
and those of $Y_0$ are
$$ \Psi_x = - \de u_x \ , \ \Psi_t = - (1 + \de ) c_0 t^\de u_x - u_t \ , \ \Psi_{xx} = - 2 \de u_{xx} \ . $$
Using these, the theorem follows by explicit computations. \EOP

It turns out that reduction, and invariant solutions, under the
vector field $X_0$ are of special interest. This is due to the
following theorem, which provides an analytic explanation of the
numerically observed behavior.

\medskip\noindent
{\bf Theorem 4.} {\it The equation (5.2) admits an asymptotically
$X_0$-invariant solution, described by (5.6).}

\medskip\noindent
{\bf Proof.} In this case the symmetry adapted coordinates are
$$ \s = t \ , \ y = (x / t^\de ) - c_0 t \ , \ w = u \ ; $$
the relevant $u$ derivatives are expressed in the new coordinates as
$$ u_t = w_\s - \( { \de y + c_0 (1 + \de ) \s \over \s } \) w_y \ , \ u_x = {1 \over t^\de} w_y \ , \ u_{xx} = {1 \over t^{2 \de}} w_{yy} \ .  $$

In these coordinates the equation (5.2) is written as
$$ w_\s \ = \ A \, w_{yy} \, + \, B \, w_y \, + \, f(w) \ , \eqno(6.4) $$
with
$$ A \ = \ \( {y+c_0 \s \over \s}\right)^{2-\a } \ \ ; \ \  B \ = \ \( {c_0 + \de {y + c_0 \s \over \s} + \epsilon \( {y + c_0 \s \over \s } \)^{2-\a} {1 \over y+c_0 \s}} \) \ . $$
The vector field $X_0$ reads simply $ X_0  = \s \pa_\s$; its
second prolongation is
$$ Y_0 \ = \ \s {\pa \over \pa \s} - w_\s {\pa \over \pa w_\s} - w_{\s y} {\pa \over \pa w_{\s y} } - 2 w_{\s \s} {\pa \over \pa w_{\s \s} } \ .  $$

For $\s \to \infty$, (6.4) reads
$$ w_\s \ = \ c_0^{2-\a} \, w_{yy} \ + \ c_0 (1+\de) \, w_y \ + \ f(w) \ . $$
The $X_0$-invariant solutions satisfy $w_\s=0$, and are thus
obtained as solution to
$$ c_0^{2-\a} w_{yy} \ + \ c_0 (1+\de) w_y \ + \ w \ =\ 0 \ , \eqno(6.5) $$
where we have used $w << 1$ in the region we are investigating (i.e. for $\s \to \infty$), so that $f(w) \simeq w$.

Solutions to (6.5) are of the form
$$ w (y) \ = \ c_1 e^{-\om_+ y} \ + \ c_2 e^{-\om_- y} \ ,  $$
where
$$ \om_\pm = { (1 +  \de) \over 2 c_0^{1-\a} } \[ {1 \pm \sqrt{1-{4 \over c_0^\a (1+\de)^2}} } \] \ . $$
If we require the solutions to be non oscillating, this implies a lower bound on the parameter $c_0$, i.e. $c_0 \ge (2/(1+\de))^{2 /\a}$.

The solution $e^{-\om_+ z}$ is unstable against small
perturbations, while $e^{-\omega_- z}$ is stable \cite{KPP}. As
proved by Kolmogorov, the asymptotic solution is the stable one
with the lowest speed giving nonoscillating behavior, i.e. $ c_0 =
[ 2 / (1 + \de) ]^{2 /\a}$. This means $ w(y) \simeq e^{- \om_0 y
}$ with $\om_0 = [2 / (1+\de)]^{1-2/\a}$. Going back to the
original variables, we get
$$ u(x,t) \ \simeq \ A \ \exp{\[ {-{x-v(t) t \over \la (t)}}\]} \ = \
A \ \exp \[ - \om_0 {{x-c_0 t^{1+\de} \over t^\de}} \] \ . $$ This
is precisely the numerically observed asymptotic behavior (5.6).
\EOP

\section{Conclusions and discussion}

We provided suitable definitions of asymptotic symmetries -- both
in proper and in conditional or partial sense -- and proposed a
method for the analysis of asymptotic symmetry properties of PDEs
and their solutions.

We applied our general method to a model class of anomalous
reaction-diffusion (ARD) equations, discussed and studied
numerically in \cite{MVV2}. We first considered the standard FKPP
equation, and described in detail its asymptotic symmetry
properties; we have also shown that our approach recovers the well
known asymptotic properties of FKPP solutions.

We have then tackled general ARD equations in our model class,
i.e. with anomalous diffusion associated to (3.5). We recalled the
features of asymptotic solutions as observed in numerical
experiments, and identified the Lie generator $X$ of the observed
asymptotic generalized scaling invariance; in theorem 1 we showed
that this is {\it not} a symmetry, but it is an asymptotic
symmetry, of the ARD equation. We have then considered the
solution-preserving maps associated to this asymptotic scaling
symmetry, focusing on the physical value of the parameter $\de$;
in theorem 2 we have shown that in this case the ARD equation has
no solution invariant under $X$ (which therefore is not a
conditional symmetry of the equation), but admits solutions which
are {\it asymptotically} invariant under it. That is, $X$ is an
asymptotic conditional symmetry of the ARD equation.

Finally, in sect.6 we passed to consider in more detail the
numerically observed  asymptotic solution (5.6). We identified the
full symmetry algebra $\G$ of it; vector fields in this algebra
are asymptotic conditional symmetries of the ARD equation. We
focused in particular on certain vector fields in $\G$, and shown
they are partial symmetries for the ARD equation. Among these
vector field is the scaling vector field $X_0$ given by (6.2) and
not depending, nor acting, on the dependent variable $u$. We
proved that the ARD equation does admit an asymptotically
$X_0$-invariant solution, which is precisely the numerically
observed one (5.6).

We have thus provided an analytic explanation for the numerically
observed behavior, based on our general method.
\bigskip

Let us conclude by presenting some brief final remarks on our
method.

{\tt (a)} This method represents an evolution of the classical
method to determine partially invariant solutions for symmetric
PDEs \cite{Olv1,Win}, and a blend of it with the method of
conditional and partial symmetries \cite{CGpar}, in order to
analyze equations which do not have complete (as opposed to
asymptotic) symmetries. Our method is based on the abstract
approach developed in \cite{Gae}, based itself on ideas and
previous work by several authors \cite{Bar,BK,CE,Gold}.

{\tt (b)} The application of the method to (generalized) scaling
symmetries and our model class of ARD equations was greatly
facilitated by the form of the vector fields and of the initial
equations $\De_0$: indeed, the $W$-accessible part of $\F$ was
finite-dimensional.

{\tt (c)} Our method deals, strictly speaking, with properties
which are asymptotic in the group parameter; these correspond to
properties asymptotic in time and space only if the considered
vector field has favourable properties itself. Also, our method
does not intend to tackle intermediate asymptotics \cite{Bar}.

{\tt (d)} This work was concerned only with scale invariance (at
infinity or near a travelling front). We trust however that
suitable generalizations of our approach can also deal with more
general asymptotic invariance properties, and more general
differential equations, and that it is potentially capable to
provide a sound explanation -- or prediction -- of the asymptotic
invariance of their solutions.

\bigskip\bigskip


\end{document}